\newcommand{\AJMRequirePackage}{}
\let\AJMRequirePackage=\RequirePackage
\def\RequirePackage#1{\typeout{%
  Loading of package `#1' deferred (don't forget it in defs.sty)}}
\documentclass[copyright,creativecommons]{eptcs}
\let\RequirePackage=\AJMRequirePackage
\let\usepackage=\RequirePackage
\providecommand{\event}{LFMTP 2011} 
\usepackage{defs}

\title{An Improved Implementation and \titlebreak
       Abstract Interface for Hybrid}
\author{Alan J. Martin
\institute{Department of Mathematics and Statistics,\\
University of Ottawa, Canada}
\email{amart045@site.uottawa.ca}
\and
Amy P. Felty
\institute{School of Electrical Engineering and Computer Science and\\
Department of Mathematics and Statistics,\\
University of Ottawa, Canada}
\email{afelty@site.uottawa.ca}
}

\def\titlerunning{An Improved Implementation and Abstract Interface
  for Hybrid}
\def\authorrunning{Alan J. Martin \& Amy P. Felty}

\begin{document}
\maketitle

\begin{abstract}
Hybrid is a formal theory implemented in Isabelle\slash HOL that provides
an interface for representing and reasoning about object languages using
higher-order abstract syntax (HOAS).  This interface is built around an
HOAS variable-binding operator that is constructed definitionally from
a de~Bruijn index representation.  In this paper we make a variety of
improvements to Hybrid, culminating in an abstract interface that on one
hand makes Hybrid a more mathematically satisfactory theory, and on the
other hand has important practical benefits.  We start with a modification
of Hybrid's type of terms that better hides its implementation in terms
of de~Bruijn indices, by excluding at the type level terms with dangling
indices.  We present an improved set of definitions, and a series of new
lemmas that provide a complete characterization of Hybrid's primitives
in terms of properties stated at the HOAS level.  Benefits of this new
package include a new proof of adequacy and improvements to reasoning
about object logics.  Such proofs are carried out at the higher level
with no involvement of the lower level de~Bruijn syntax.
\end{abstract}

\section{Introduction}

Hybrid is a system developed to specify and reason about logics,
programming languages, and other formal systems expressed in
higher-order abstract syntax (HOAS).  It is implemented as a formal
theory in Isabelle/HOL~\cite{nipkow/paulson/wenzel:2002}.  By
providing HOAS in a modern proof assistant, Hybrid automatically gains
the latter's capabilities for meta-theoretical reasoning.  This
approach is intended to provide advantages in flexibility and proof
automation, in contrast to systems that directly implement logical
frameworks, which must build their own meta-reasoning layers from the
ground up.  Building a system such as Hybrid within a general purpose
theorem prover poses a variety of challenges.  Our goal in this work
is to improve the implementation and interface of Hybrid's basic
theory, bringing it to a point where its potential advantages can be
more fully realized.

Using HOAS, binding constructs in the represented language (the
\emph{object logic} or OL) are encoded using the binding constructs
provided by an underlying $\lambda$-calculus or function space of the
meta-logic, thus representing the arguments of these constructs as
functions of the meta-level.  Isabelle/HOL implements an extension of
higher-order logic, where the function types are ``too large'' for
HOAS in two senses.  First, they contain elements with irreducible
occurrences of logical constants, which do not represent syntax.
Second, the function space $\tau \Rightarrow \tau$ has larger
cardinality than $\tau$, so a variable-binding operator represented as
a functional $\Phi$ of type $(\tau \Rightarrow \tau)\Rightarrow \tau$
cannot be injective.  This makes it unsuitable for syntax, for we
cannot uniquely recover the argument $F$ from a term of the form
$\Phi(F)$.  Our work builds directly on the original Hybrid
system~\cite{ambler/crole/momigliano:2002}, whose solution to both
problems is to use only a \emph{subset} of the funtion type,
identified by a predicate called \(\notationII \definedterm{abstr}\).  It builds a type
\(\notationII \isaholtype{expr}\) of terms with an HOAS variable-binding operator
\emph{definitionally} in terms of a de Bruijn index representation.

In earlier work joint with Alberto Momigliano, we gave a system
presentation of Hybrid~\cite{momigliano/martin/felty:2008}, which
built on the original Hybrid and serves as a starting point for the
work presented here.  In this paper, we fill in many details that
could not be described in a short system description, as well as make
significant further improvements, allowing us to complete a
characterization of Hybrid's type \(\notationII \isaholtype{expr}\) in terms of properties
stated at the HOAS level.  In the new Hybrid, the type \(\notationII \isaholtype{expr}\), its
constructors, and these properties form an abstract interface that
allows users to reason at the higher level with no involvement of the
lower level implementation details.  This interface was motivated by
and is illustrated by a new proof of representational adequacy for
Hybrid~\cite[Sect.~3.4]{martin:2010a} that does not make any reference
to de Bruijn syntax.

We start in \abbrevref{sec:hybrid-overview} by giving an abstract view
of Hybrid that motivates and explains the interface.
Sections~\ref{sec:hybrid-defn-deBruijn}--\ref{sec:hybrid-abstr2-and-abstr-LAM}
fill in many of the details of its implementation.
The type \(\notationII \isaholtype{dB}\) implementing the de Bruijn index representation is
defined in \abbrevref{sec:hybrid-defn-deBruijn}, along with a
predicate \(\notationII \definedterm{level}\) to keep track of dangling indices.  The original
Hybrid~\cite{ambler/crole/momigliano:2002} used a datatype
corresponding to our \(\notationII \isaholtype{dB}\) directly as \(\notationII \isaholtype{expr}\).
\abbrevref{sec:hybrid-defn-expr} defines the new version of
\(\notationII \isaholtype{expr}\), which excludes at the type level terms with dangling
indices. This simplifies the representation of object languages by
eliminating the need to carry a predicate for this purpose (called
\(\notationII \definedterm{proper} \) in~\cite{ambler/crole/momigliano:2002}) along with
Hybrid terms in meta-theoretic reasoning.
Section~\ref{sec:hybrid-defn-abstr-LAM} defines Hybrid's variable
binding operator \(\notationII \definedterm{LAM}\) and the \(\notationII \definedterm{abstr}\) predicate.  These
definitions support a stronger injectivity property, presented in
Sect.~\ref{sec:hybrid-defn-LAM-inject} with only one \(\notationII \definedterm{abstr}\)
premise rather than two.  This property was also proved
in~\cite{momigliano/martin/felty:2008}; the results here generalize
and simplify these definitions as well as simplify other related
Hybrid internals.  (In particular, we eliminate the need for the
auxiliary function \(\notationII \isaholtype{dB\_fn}\) defined
in~\cite{momigliano/martin/felty:2008} using the function package first
introduced in Isabelle/HOL 2007, and we eliminate some other auxiliary
functions by using a more systematic treatment of \(\notationII \definedterm{level}\).)

In \abbrevref{sec:hybrid-abstr2-and-abstr-LAM}, we formally prove that
a version of \(\notationII \definedterm{abstr}\) for two-argument functions (as described
in~\cite{momigliano/ambler/crole:2002}) is equivalent to a conjunction
of one-argument \(\notationII \definedterm{abstr}\) conditions on ``slices'' of the function
(fixing one argument).  We use this result to prove a case-distinction
lemma for functions satisfying \(\notationII \definedterm{abstr}\), and a lemma that enables
compositional proof of \(\notationII \definedterm{abstr}\) conditions at the HOAS level,
without conversion to de Bruijn indices as required
in~\cite{ambler/crole/momigliano:2002}.  These two lemmas represent
important new results that complete the abstract interface for Hybrid.

In \abbrevref{sec:conclude}, we discuss related work as well as
ongoing work with Hybrid.

The Isabelle/HOL 2011 theory file for the present version of Hybrid is
available online at:
\begin{center}
\url{http://hybrid.dsi.unimi.it/download/Hybrid.thy}
\end{center}
and a more thorough presentation can be found in the first author's
Ph.D.\ thesis~\cite{martin:2010a,martin:2010b}.
In addition to the results described here, this theory file also
replaces tactic-style proofs of the original version of Hybrid with
Isar proofs.  This style of proof is both more readable and more
robust against changes to the underlying proof assistant.  It also
includes rewrite rules for Isabelle's simplifier to convert
automatically between HOAS at type \(\notationII \isaholtype{expr}\) and de Bruijn indices at
type \(\notationII \isaholtype{dB}\).  With the improvements allowing users to work
exclusively at the HOAS level, this is no longer needed, and only
included for illustrative purposes.

\section{An Abstract View of Hybrid}
\label{sec:hybrid-overview}

We use a pretty-printed version of Isabelle\slash HOL concrete syntax
in this and the following sections.
A double colon \(\notationII  \isaoftype \) separates a term from its type, and the arrow
\(\notationII  \isatypearrow \) is used in function types.
We stick to the usual logical symbols for connectives and quantifiers
(\(\notationII  \neg \), \(\notationII  \isaholconj \), \(\notationII  \isaholdisj \), \(\notationII  \isaholimpl \), \(\notationII   \isaholuniv \), \(\notationII   \isaholexists \)).
Free variables (upper-case) are implicitly universally quantified
(from the outside).
The sign \(\notationII  \isaequiv \)\ (Isabelle meta-equality) is used for equality by
definition, and \(\notationII  \isaimpl \) for Isabelle meta-level implication.
In the notation \(\notationII  \isabigldblbrack P_1; \ldots; P_n \isabigrdblbrack \isaimpl P \), the square
brackets are used to group premises to abbreviate nested implications;
in its expanded form, it is \(\notationII  P_1 \isaimpl \dots \isaimpl P_n \isaimpl P \).
Similarly, \(\notationII  \isabiglbrack t_1, \ldots, t_n \isabigrbrack \isatypearrow t \) abbreviates the type
\(\notationII  t_1 \isatypearrow \dots \isatypearrow t_n \isatypearrow t \).
The keyword \(\notationII  \isacommand{datatype} \) introduces a new datatype, while
\(\notationII  \isacommand{function} \) introduces a recursively defined function.
We freely use infix notations, often without explicit declarations.
Other syntax is intrduced as it appears.

Isabelle\slash HOL already has extensive support for first-order abstract
syntax, in the form of its \(\notationII \isacommand{datatype}\) package.  Hybrid may be viewed
as an attempt to approximate a \(\notationII \isacommand{datatype}\) definition that is not
well-formed because of its higher-order features:
\begin{Display} \formal
\(\notationII   \isacommand{datatype}\ \isaholtype{expr} = \definedterm{CON}\ \isaholtype{con}\ {\big|}\ \definedterm{VAR}\ \isaholtype{var} %
  \ {\big|}\ \definedterm{APP}\ \isaholtype{expr}\ \isaholtype{expr} \quad (\text{notation } (\localvar{s} \hybridapp \localvar{t})) \)\par\nopagebreak
\(\notationII   \phantom{ \isacommand{datatype}\ \isaholtype{expr} = } \! %
      {\big|}\ \definedterm{LAM}\ (\isaholtype{\underline{expr}} \isatypearrow \isaholtype{expr}) %
    \quad (\text{notation } (\definedterm{LAM}\ \localvar{x}. \localvar{B})) \)
\end{Display}
where \(\notationII \definedterm{CON}\) represents constants, from an OL-specific type \(\notationII \isaholtype{con}\)
(typically a trivial \(\notationII \isacommand{datatype}\)); \(\notationII \definedterm{VAR}\) may be used to represent
free variables, from a countably infinite type \(\notationII \isaholtype{var}\)\ (actually
a synonym for \(\notationII \isaholtype{nat}\)); \(\notationII \definedterm{APP}\) represents pairing, which is sufficient
to encode list- or tree-structured syntax; and \(\notationII \definedterm{LAM}\) represents variable
binding in HOAS style, using the bound variable of an Isabelle\slash HOL
$\lambda$-abstraction to represent a bound variable of the object language.%
  \footnote{%
    While \(\notationII \definedterm{APP}\) and \(\notationII \definedterm{LAM}\) were inspired by the untyped $\lambda$-%
    calculus, in Hybrid they are used only as syntax, without built-in notions
    of $\beta$-conversion, normal forms, etc.}

It should be noted that Hybrid only approximates \emph{one} such
pseudo-datatype, not the \(\notationII \isacommand{datatype}\) package with its ability
to define multiple types for first-order abstract syntax.  That is,
Hybrid is \emph{untyped}, so predicates rather than types must be used
to distinguish different kinds of OL terms encoded into \(\notationII \isaholtype{expr}\).

The problem with the above definition
is \(\notationII \definedterm{LAM}\), whose argument type includes a negative occurrence
of \(\notationII \isaholtype{expr}\)\ (underlined above).  This is essential for HOAS, but it is not
permitted in a \(\notationII \isacommand{datatype}\) definition
\cite[Sect.~2.6]{nipkow/paulson/wenzel:2011}, and it will require modifications
to some of the properties expected for a constructor of a datatype;
we will return to this issue later.

Hybrid does provide a type \(\notationII \isaholtype{expr}\) with operators \(\notationII \definedterm{CON}\), \(\notationII \definedterm{VAR}\),
\(\notationII \definedterm{APP}\), and \(\notationII \definedterm{LAM}\) of the appropriate types.  This type and the latter
three operators can be used directly as a representation of the untyped
$\lambda$-calculus.
When encoding OLs in general, however, it is usual to represent each OL
construct as a list built using \(\notationII  \hybridapp \) and headed by a \(\notationII \definedterm{CON}\) term
identifying the particular construct.  
To illustrate this idea, we take the untyped $\lambda$-calculus as our
OL with its usual named-variable syntax, using capital letters for
variables (\(\notationI  V_i \), \(\notationI  i \in \mathbb{N} \)) and
$\lambda$-abstraction (\(\notationI \Lambda\)) to avoid confusion with
Isabelle's $\lambda$ operator.  In this form, an object language term
  \(\notationI  (\Lambda\ V_1 \ldotp \Lambda\ V_2 \ldotp (V_1 \  V_2) \  V_3) \),
for example, can be represented as
  \[\notationII  \definedterm{c\_lam} \hybridapp (\definedterm{LAM}\ \localvar{x}. \definedterm{c\_lam} \hybridapp (\definedterm{LAM}\ \localvar{y}. %
       \definedterm{c\_app} \hybridapp (\definedterm{c\_app} \hybridapp \localvar{x} \hybridapp \localvar{y}) \hybridapp \definedterm{VAR}\ 3)), \]
where \(\notationII  \definedterm{c\_lam} = \definedterm{CON}\ c_1 \) and \(\notationII  \definedterm{c\_app} = \definedterm{CON}\ c_2 \) for
distinct constants \(\notationIII  c_1, c_2 :: \isaholtype{con} \).  We may use Isabelle's ability
to define abbreviations and infix notations to recover a reasonable concrete
syntax:
  \[\notationII  \definedterm{fn}\ \localvar{x}\ \localvar{y}. (\localvar{x} \objectapp \localvar{y}) \objectapp \definedterm{VAR}\ 3.\]
Note that although de Bruijn indices do not appear in such terms, numbers
can appear as arguments to Hybrid's \(\notationII \definedterm{VAR}\) operator, which is
included to allow a representation of free variables that is distinct
from bound variables.

We now turn to the properties required of \(\notationII \isaholtype{expr}\) and its operators
to function as HOAS.  We motivate the
requirements by considering adequacy, an important meta-theoretic
property.  This can take several forms, but the proof presented
in~\cite{martin:2010a} uses bijectivity of a set-theoretic semantics on
a $\lambda$-calculus-like subset of the Isabelle\slash HOL terms of
type \(\notationII \isaholtype{expr}\), called the \defterm{syntactic terms}:
  \[\notationIII  s ::= x \mid \definedterm{CON}\ a \mid \definedterm{VAR}\ n \mid s_1 \hybridapp s_2 \mid \definedterm{LAM}\ x \ldotp s \]
where \(\notationIII s\) (with possible subscripts) stands for a syntactic term,
\(\notationIII x\) for a variable of type \(\notationIII \isaholtype{expr}\), \(\notationIII a\) for a constant of
type \(\notationIII \isaholtype{con}\), and \(\notationIII n\) for a natural-number constant.
Note that $s$ is an informal mathematically defined set; it is not a
formal Isabelle\slash HOL definition.

However, open terms present a complication.  Suppose we have a theory where
the semantics is bijective on \emph{closed} syntactic terms, which it maps
to a set \(\notationIII S\).  Then it will map \emph{open} terms with \(\notationIII n\) free variables
to functions from the Cartesian power \(\notationIII S^n\) to \(\notationIII S\).  But there are many
such functions that do not correspond to syntactic terms; for example, the
function \(\notationIII  S \rightarrow S \) corresponding to the Isabelle\slash HOL term
  \[\notationII  \isalambda\ \localvar{x}. \definedterm{if}\ ( \isaholexists\ \localvar{a}. \localvar{x} = \definedterm{CON}\ \localvar{a})\ \definedterm{then}\ (\localvar{x} \hybridapp \localvar{x})\ \definedterm{else}\ \localvar{x} \]
of type \(\notationIII  (\isaholtype{expr} \isatypearrow \isaholtype{expr}) \).  Indeed, there are a countable infinity of
syntactic terms, while the set of functions from \(\notationIII S^n\) to \(\notationIII S\) is
uncountable for \(\notationIII  n \geq 1 \).

Thus, Hybrid must define a \emph{subset} of the function space to be used as
its representation for open syntactic terms.  This is done using a predicate
\(\notationII  \definedterm{abstr} \isaoftype ((\isaholtype{expr} \isatypearrow \isaholtype{expr}) \isatypearrow \isaholtype{bool}) \).  The functions satisfying
\(\notationII \definedterm{abstr}\) will be those of the form
\(\notationIII  (\isalambda  \localvar{x} \ldotp s) \) where \(\notationIII s\) is
a syntactic term with (at most) one free variable \(\notationII \localvar{x}\); we call these the
\emph{syntactic functions}.%
  \footnote{%
    Previous work called such functions \emph{abstractions}
    \cite{ambler/crole/momigliano:2002}~-- thus the predicate name
    \(\notationII \definedterm{abstr}\); and called functions not satisfying \(\notationII \definedterm{abstr}\)
    \emph{exotic terms} \cite{ambler/crole/momigliano:2002,%
    despeyroux/felty/hirschowitz:1995}.
}
(Syntactic terms with more than one free variable can be handled one variable
at a time.)

In the first-order case, three properties hold of a type defined using
Isabelle\slash HOL's \(\notationII \isacommand{datatype}\): \emph{distinctness} of the
datatype constructors, \emph{injectivity} of each constructor, and an
\emph{induction principle}.
In the case of Hybrid, distinctness of all the operators and injectivity of
the first-order operators (i.e., all except \(\notationII \definedterm{LAM}\)) are straightforward
to achieve, e.g.:
\begin{Display} \formal
\(\notationII   \isaholuniv\ (\localvar{c} \isaoftype \isaholtype{con})\ (\localvar{S} \isaoftype \isaholtype{expr} \isatypearrow \isaholtype{expr}). \definedterm{CON}\ \localvar{c} \neq \definedterm{LAM}\ \localvar{S} \)\par\nopagebreak
\(\notationII   \isaholuniv\ (\localvar{s}\ \localvar{t}\ \localvar{s'}\ \localvar{t'}:: \isaholtype{expr}). (\localvar{s} \hybridapp \localvar{t} = \localvar{s'} \hybridapp \localvar{t'}) \isaholimpl %
     (\localvar{s} = \localvar{s'}) \isaholconj (\localvar{t} = \localvar{t'}) \ldotp \)
\end{Display}
(These properties are used as rewrite rules for Isabelle's simplifier, to
reduce equalities of Hybrid terms with known operators on both sides; typically
this results in equalities where one side is just an Isabelle\slash HOL
variable, which can then be eliminated by substitution.%
  \footnote{%
    Indeed, most use of Hybrid's lemmas in object-language work is automated
    using Isabelle's simplifier and classical reasoner, and as a result,
    direct references to Hybrid's lemmas may be rare.})

Injectivity of \(\notationII \definedterm{LAM}\) must be restricted to functions satisfying
\(\notationII \definedterm{abstr}\); indeed, it can be proven in Isabelle\slash HOL that no
\vadjust{\pagebreak[3]}%
injective function from \(\notationII  (\isaholtype{expr} \isatypearrow \isaholtype{expr}) \) to \(\notationII \isaholtype{expr}\) exists,
by formalizing Cantor's diagonal argument.  As mentioned earlier, our
improved version requires an \(\notationII \definedterm{abstr}\)
condition for only one side of the equality:
  \[\notationII  \isabigldblbrack \definedterm{abstr}\ \localvar{S} \isaholdisj \definedterm{abstr}\ \localvar{T}; \definedterm{LAM}\ \localvar{S} = \definedterm{LAM}\ \localvar{T} \isabigrdblbrack \isaimpl \localvar{S} = \localvar{T} \ldotp \]
Requiring only a single condition
reduces the need for explicit \(\notationII \definedterm{abstr}\) conditions in object-language
encodings, because they can be transported across equalities of \(\notationII \definedterm{LAM}\)
terms.  It is achieved by adding to the type \(\notationII \isaholtype{expr}\) an additional
constant \(\notationII \definedterm{ERR}\), and defining \(\notationII \definedterm{LAM}\) to take the value \(\notationII \definedterm{ERR}\) on
functions not satisfying \(\notationII \definedterm{abstr}\).  (The constant \(\notationII \definedterm{ERR}\) will sometimes
appear as an additional case alongside the operators of Hybrid, in lemmas
that impose an \(\notationII \definedterm{abstr}\) condition for the \(\notationII \definedterm{LAM}\) case.  We also include
it among the syntactic terms.)

Since \(\notationII \definedterm{abstr}\) appears as a premise of injectivity---and it would in any
case be needed to state properties of open syntactic terms---we must also
include properties sufficient to characterize it.  While Hybrid proves a
number of lemmas regarding \(\notationII \definedterm{abstr}\) for convenience and proof automation,
the desired characterization can be given in a single statement:
{\setbox0=\hbox{\(\notationII  ( \isaholexists\ \localvar{S}\ \localvar{T}. {} \)}
 \xdef\ajmtemp{\the\wd0}}
\begin{Display} \formal
\(\notationII  \definedterm{abstr}\ \localvar{Y} \isaequiv %
   \quad \phantom{ \isaholdisj } \quad
     \makebox[\ajmtemp][r]{\(\notationII  ( \)}
       \localvar{Y} = (\isalambda\ \localvar{x}. \localvar{x})) \)\par\nopagebreak
\(\notationII  \phantom{\definedterm{abstr}\ \localvar{Y} \isaequiv {}} \quad { \isaholdisj } \quad
     \makebox[\ajmtemp][r]{\(\notationII  ( \isaholexists\ \localvar{a}. {} \)}
       \localvar{Y} = (\isalambda\ \localvar{x}. \definedterm{CON}\ \localvar{a})) \)\par\nopagebreak
\(\notationII  \phantom{\definedterm{abstr}\ \localvar{Y} \isaequiv {}} \quad { \isaholdisj } \quad
     \makebox[\ajmtemp][r]{\(\notationII  ( \isaholexists\ \localvar{n}. {} \)}
       \localvar{Y} = (\isalambda\ \localvar{x}. \definedterm{VAR}\ \localvar{n})) \)\par\nopagebreak
\(\notationII  \phantom{\definedterm{abstr}\ \localvar{Y} \isaequiv {}} \quad { \isaholdisj } \quad
     \makebox[\ajmtemp][r]{\(\notationII  ( \isaholexists\ \localvar{S}\ \localvar{T}. {} \)}
       \localvar{Y} = (\isalambda\ \localvar{x}. \localvar{S}\ \localvar{x} \hybridapp \localvar{T}\ \localvar{x}) \isaholconj \definedterm{abstr}\ \localvar{S} \isaholconj \definedterm{abstr}\ \localvar{T}) \)\par\nopagebreak
\(\notationII  \phantom{\definedterm{abstr}\ \localvar{Y} \isaequiv {}} \quad { \isaholdisj } \quad
     \makebox[\ajmtemp][r]{\(\notationII  ( \isaholexists\ \localvar{W}. {} \)}
       \localvar{Y} = (\isalambda\ \localvar{x}. \definedterm{LAM}\ \localvar{y}. \localvar{W}\ \localvar{x}\ \localvar{y}) \isaholconj \underline{\definedterm{abstr}\ \localvar{W}}) \)\par\nopagebreak
\(\notationII  \phantom{\definedterm{abstr}\ \localvar{Y} \isaequiv {}} \quad { \isaholdisj } \quad
     \makebox[\ajmtemp][r]{\(\notationII  ( \)}
       \localvar{Y} = (\isalambda\ \localvar{x}. \definedterm{ERR})) \)
\end{Display}

Once again the \(\notationII \definedterm{LAM}\) case complicates matters: the underlined occurrence
of \(\notationII  (\definedterm{abstr}\ \localvar{W}) \) applies \(\notationII \definedterm{abstr}\) to a function
  \(\notationII  \localvar{W} \isaoftype \bigl( {[} \isaholtype{expr}, \isaholtype{expr} {]} \isatypearrow \isaholtype{expr} \bigr) \).
This should be possible by using type classes to give a polymorphic definition
for \(\notationII \definedterm{abstr}\), but that is future work.  The present version of Hybrid
instead replaces \(\notationII  (\definedterm{abstr}\ \localvar{W}) \) with
  \(\notationII  ( \isaholuniv\ \localvar{y}. \definedterm{abstr}\ (\isalambda\ \localvar{x}. \localvar{W}\ \localvar{x}\ \localvar{y})) \isaholconj %
     ( \isaholuniv\ \localvar{x}. \definedterm{abstr}\ (\isalambda\ \localvar{y}. \localvar{W}\ \localvar{x}\ \localvar{y})) \).

As for induction, it can take several forms.  First, a kind of size
induction on \(\notationII \isaholtype{expr}\) is available, similar to size induction for
types defined by Isabelle\slash HOL's datatype package.  This
induction has limited applicability in the higher-order setting,
although it was used in the proof of adequacy~\cite{martin:2010a}.
We also retain an induction principle from the original version of
Hybrid~\cite{ambler/crole/momigliano:2002} where the first-order
induction cases are standard, while the \(\notationII \definedterm{LAM}\) case is:
\[\notationII   \isaholuniv\ \localvar{S} \isaoftype (\isaholtype{expr} \isatypearrow \isaholtype{expr}){.}\quad %
   \definedterm{abstr}\ \localvar{S} \isaholconj \bigl(  \isaholuniv\ \localvar{n}. \localvar{P}\ (\localvar{S}\ (\definedterm{VAR}\ \localvar{n})) \bigr) %
     \isaholimpl \localvar{P}\ (\definedterm{LAM}\ \localvar{x}. \localvar{S}\ \localvar{x}) \ldotp \]

A common form of induction used in many case studies involves some
form of structural induction on the encoding of the inference rules of
an OL.  For this kind of reasoning, a \emph{two-level} approach is
adopted, similar in spirit to other systems such as
\emph{Twelf}~\cite{pfenning/schurmann:1999} and
\emph{Abella}~\cite{gacek:2008}.  An intermediate layer between the
meta-logic (Isabelle/HOL) and the OL, called a \emph{specification
  logic}, is defined inductively in Isabelle/HOL.  This middle layer
allows succinct and direct encodings of object logic inference rules,
which are also defined as inductive definitions.  
Successful applications of this kind of induction can be found
in~\cite{felty/momigliano:2008,martin:2010a}, for example.

Finally, Hybrid aims to build \(\notationII \isaholtype{expr}\) and its operators definitionally
in Isabelle\slash HOL.  While the description above is an informal but
reasonably complete specification of Hybrid, it is not directly usable
as a definition because it is circular: the arguments of \(\notationII \definedterm{LAM}\) and
\(\notationII \definedterm{abstr}\) may themselves contain \(\notationII \definedterm{LAM}\), and injectivity of \(\notationII \definedterm{LAM}\)
depends on \(\notationII \definedterm{abstr}\).  It could be formalized as an axiomatic theory,
leaving consistency as a meta-theoretical problem; but instead, Hybrid is
built definitionally in terms of a \emph{first-order} representation of
variable binding based on de~Bruijn indices.  The definitions and lemmas
involved in achieving this are the subject of the next sections.

\section{De Bruijn syntax}
\label{sec:hybrid-defn-deBruijn}

The Hybrid theory defines the type \(\notationII \isaholtype{expr}\) in terms of an Isabelle\slash
HOL datatype \(\notationII \isaholtype{dB}\), which represents abstract syntax using a nameless
first-order representation of bound variables called \defterm{de~Bruijn
indices} \cite{debruijn:1972}.

This approach differs from the original version of Hybrid
\cite{ambler/crole/momigliano:2002}, which used a datatype corresponding
to our \(\notationII \isaholtype{dB}\) directly as \(\notationII \isaholtype{expr}\); the significance of this difference
will be explained in Sections \ref{sec:hybrid-defn-expr}~and~%
\ref{sec:hybrid-defn-LAM-inject}.  However, the datatype itself is very
similar, and this section follows \cite{ambler/crole/momigliano:2002} closely.

\begin{Defn}
\label{defn:hybrid-dB-type} \formal
\(\notationII  \negquad \isacommand{types} \)\par\nopagebreak
\(\notationII    \isaholtype{var} = \isaholtype{nat} \)\par\nopagebreak 
\(\notationII    \isaholtype{bnd} = \isaholtype{nat} \)\par\nopagebreak[3]
\(\notationII  \negquad \isacommand{datatype}\ \isatypevar{a}\ \isaholtype{dB} = \vphantom{ {\big|} } \)\par\nopagebreak
\(\notationII  \phantom{ {\big|} } %
  \ \definedterm{CON'}\ \isatypevar{a}\ {\big|}\ \definedterm{VAR'}\ \isaholtype{var}\ {\big|}\ \definedterm{APP'}\ (\isatypevar{a}\ \isaholtype{dB})\ (\isatypevar{a}\ \isaholtype{dB}) %
     \quad (\text{notation } (\localvar{s} \hybridappO \localvar{t})) \)\par\nopagebreak
\(\notationII  {\big|}\ \definedterm{ERR'}\ {\big|}\ \definedterm{BND'}\ \isaholtype{bnd}\ {\big|}\ \definedterm{ABS'}\ (\isatypevar{a}\ \isaholtype{dB}) \)
\end{Defn}

The constructors \(\notationII \definedterm{CON'}\), \(\notationII \definedterm{VAR'}\), and \(\notationII \definedterm{APP'}\) correspond to the
operators \(\notationII \definedterm{CON}\), \(\notationII \definedterm{VAR}\), and \(\notationII \definedterm{APP}\) on type \(\notationII \isaholtype{expr}\), which
were discussed in \abbrevref{sec:hybrid-overview} and will be defined later.
The one significant difference is that the argument of \(\notationII \definedterm{CON'}\) is a
type parameter \(\notationII \isatypevar{a}\), rather than a particular type \(\notationII \isaholtype{con}\).  This
will actually be true for \(\notationII \definedterm{CON}\) as well, and it allows Hybrid to be
defined as an OL-independent Isabelle\slash HOL theory, and later used
with OL-specific constants.  (We will frequently omit this type parameter,
except where it occurs in formal definitions or it is instantiated.)

The other three constructors (\(\notationII \definedterm{ERR'}\), \(\notationII \definedterm{BND'}\), and \(\notationII \definedterm{ABS'}\)) will
all be used in the definition of \(\notationII \definedterm{LAM}\).  The constant \(\notationII \definedterm{ERR'}\) will
be a placeholder for \(\notationII \definedterm{LAM}\) applied to a non-syntactic function; it was
not present in \cite{ambler/crole/momigliano:2002}, and its significance
will be explained later.  The constructor \(\notationII \definedterm{ABS'}\) functions as a
nameless binder, while \(\notationII  (\definedterm{BND'}\ \localvar{i}) \) represents the variable implicitly
bound by the \(\notationII  (\localvar{i} + 1)^\text{th} \) enclosing \(\notationII \definedterm{ABS'}\) node.
If there are not enough \(\notationII \definedterm{ABS'}\) nodes, then it is called a
\defterm{dangling index}.

As an example, consider the term
  \[\notationII  \underline{\definedterm{ABS'}}\ (\definedterm{ABS'} %
    \ (\definedterm{BND'}\ 2 \hybridappO \underline{\definedterm{BND'}\ 1} \hybridappO \definedterm{BND'}\ 0) \hybridappO %
        \underline{\definedterm{BND'}\ 0}) \ldotp \]
The underlined occurrences of \(\notationII  (\definedterm{BND'}\ 1) \) and \(\notationII  (\definedterm{BND'}\ 0) \) both
refer to the variable bound by the outer \(\notationII \definedterm{ABS'}\)\ (also underlined), while
the other occurrence of \(\notationII  (\definedterm{BND'}\ 0) \) refers to the variable bound by
the inner \(\notationII \definedterm{ABS'}\).  \(\notationII  (\definedterm{BND'}\ 2) \) is a dangling index, because there
are only 2 enclosing \(\notationII \definedterm{ABS'}\) nodes.

To keep track of dangling indices, we define a predicate
  \(\notationII  \definedterm{level} \isaoftype \isabiglbrack \isaholtype{bnd}, \isaholtype{dB} \isabigrbrack \isatypearrow \isaholtype{bool} \)
such that \(\notationII  (\definedterm{level}\ \localvar{i}\ \localvar{t}) \) is true if enclosing the term \(\notationII \localvar{t}\) in
\(\notationII \localvar{i}\) or more \(\notationII \definedterm{ABS'}\) nodes would result in a term without dangling
indices.  (We omit the formal definition, which is straightforward.)
A term with no dangling indices is called \defterm{proper}, and we may
define an abbreviation
  \(\notationII  (\definedterm{proper}\ \localvar{t}) \isaholiff (\definedterm{level}\ 0\ \localvar{t}) \).
These notions are standard for abstract syntax based on de~Bruijn indices
\cite{ambler/crole/momigliano:2002}.

\section{The type \ldquo expr\rdquo\ of proper de Bruijn terms}
\label{sec:hybrid-defn-expr}

Defining a type designed specifically to represent syntax has been
used in a variety of approaches to reasoning about the
$\lambda$-calculus and other object logics
(e.g.~\cite{Norrish:HOSC2006,Urban:JAR2008}).
Here, we use Isabelle\slash HOL's
\(\notationII \isacommand{typedef}\) mechanism to define \(\notationII \isaholtype{expr}\) as a bijective image of the
set of proper terms of type \(\notationII \isaholtype{dB}\).\footnote{The version of
  \(\notationII \isaholtype{expr}\) presented here is a modification of the one used
  in~\cite{momigliano/martin/felty:2008}.}
That eliminates the \(\notationII \definedterm{proper}\)
conditions in object-language work using Hybrid, at the expense of
having to convert terms between \(\notationII \isaholtype{expr}\) and \(\notationII \isaholtype{dB}\) in defining
\(\notationII \definedterm{LAM}\) and \(\notationII \definedterm{abstr}\).  This is a good trade-off, because those
definitions are internal to Hybrid and need only be made once.  It also
turns out to be essential for strengthening the quasi-injectivity property
of \(\notationII \definedterm{LAM}\), as described in \abbrevref{sec:hybrid-defn-LAM-inject}.

\begin{Defn}
\label{defn:hybrid-expr-type} \formal
\(\notationII  \isacommand{typedef}\ (\isacommand{open}) 
  \ \isatypevar{a}\ \isaholtype{expr} = \{ \localvar{x} \isaoftype \isatypevar{a}\ \isaholtype{dB}. \definedterm{level}\ 0\ \localvar{x} \} %
  \quad \isacommand{morphisms}\ \definedterm{dB}\ \definedterm{expr} \)
\end{Defn}

This \(\notationII \isacommand{typedef}\) statement first demands a proof that the specified set
is nonempty (which is trivial here).  Then it introduces the type \(\notationII \isaholtype{expr}\),
the functions \(\notationII  \definedterm{dB} \isaoftype (\isaholtype{expr} \isatypearrow \isaholtype{dB}) \) and \(\notationII  \definedterm{expr} \isaoftype (\isaholtype{dB} \isatypearrow \isaholtype{expr}) \),
and axioms stating that they are inverse bijections between the type \(\notationII \isaholtype{expr}\)
and the set \(\notationII  \{ \localvar{x} \isaoftype \isaholtype{dB}. \definedterm{level}\ 0\ \localvar{x} \} \).
(Although axioms are used, the overall
mechanism is a form of definitional extension and preserves consistency
of the theory.)

We may now define all of the first-order operators of Hybrid (i.e., all
except \(\notationII \definedterm{LAM}\), with its functional-type argument) in the obvious way.

\begin{Defn}
\label{defn:hybrid-fo-ops} \formal
{\setbox0=\hbox{\(\notationII  \definedterm{APP} \isaoftype \isabiglbrack \isatypevar{a}\ \isaholtype{expr}, \isatypevar{a}\ \isaholtype{expr} \isabigrbrack \isatypearrow \isatypevar{a}\ \isaholtype{expr} \)\qquad}
 \xdef\ajmtemp{\the\wd0}}
\makebox[\ajmtemp][l]{\(\notationII  \definedterm{CON} \isaoftype \isatypevar{a} \isatypearrow \isatypevar{a}\ \isaholtype{expr} \)}%
\(\notationII  \definedterm{CON}\ \localvar{a} \isaequiv \definedterm{expr}\ (\definedterm{CON'}\ \localvar{a}) \)\par\nopagebreak[3]
\makebox[\ajmtemp][l]{\(\notationII  \definedterm{VAR} \isaoftype \isaholtype{var} \isatypearrow \isatypevar{a}\ \isaholtype{expr} \)}%
\(\notationII  \definedterm{VAR}\ \localvar{n} \isaequiv \definedterm{expr}\ (\definedterm{VAR'}\ \localvar{n}) \)\par\nopagebreak[3]
\makebox[\ajmtemp][l]{\(\notationII  \definedterm{APP} \isaoftype \isabiglbrack \isatypevar{a}\ \isaholtype{expr}, \isatypevar{a}\ \isaholtype{expr} \isabigrbrack \isatypearrow \isatypevar{a}\ \isaholtype{expr} \)}%
\(\notationII  \localvar{s} \hybridapp \localvar{t} \isaequiv \definedterm{expr}\ (\definedterm{dB}\ \localvar{s} \hybridappO \definedterm{dB}\ \localvar{t}) \)\par\nopagebreak
\(\notationII  \quad (\text{notation } (\localvar{s} \hybridapp \localvar{t}))\)\par\nopagebreak[3]
\makebox[\ajmtemp][l]{\(\notationII  \definedterm{ERR} \isaoftype \isatypevar{a}\ \isaholtype{expr} \)}%
\(\notationII  \definedterm{ERR} \isaequiv \definedterm{expr}\ \definedterm{ERR'} \)
\end{Defn}

\(\notationII \definedterm{ERR}\) is defined as if it were a separate operator, and it will sometimes
be treated as such, but it will also be generated by \(\notationII \definedterm{LAM}\) applied to a
non-syntactic function.

The functions \(\notationII \definedterm{dB}\) and \(\notationII \definedterm{expr}\) translate these operators to the
corresponding constructors of \(\notationII \isaholtype{dB}\)\ (\abbrevref{defn:hybrid-dB-type})
and vice versa.  This is formalized by a set of lemmas that follow
straightforwardly from the definitions, of which we present just those
for \(\notationII \definedterm{APP}\)\ (\(\notationII  \hybridapp \)) as an example.

\begin{Lem}
\label{lem:hybrid-expr-dB-simps-ex} \formal
 \(\notationII  \definedterm{dB}\ (\localvar{s} \hybridapp \localvar{t}) = \definedterm{dB}\ \localvar{s} \hybridappO \definedterm{dB}\ \localvar{t} \)\par\nopagebreak[3]
\(\notationII  \isabigldblbrack \definedterm{level}\ 0\ \localvar{s}; \definedterm{level}\ 0\ \localvar{t} \isabigrdblbrack \isaimpl %
     \definedterm{expr}\ (\localvar{s} \hybridappO \localvar{t}) = \definedterm{expr}\ \localvar{s} \hybridapp \definedterm{expr}\ \localvar{t} \)
\end{Lem}

Distinctness and injectivity for these operators follow from the corresponding
properties of \(\notationII \isaholtype{dB}\).  In \abbrevref{sec:hybrid-defn-LAM-inject}, we will
extend these results to \(\notationII \definedterm{LAM}\) as well.

The \(\notationII  (\definedterm{level}\ 0) \) premises in the lemma above are needed because the
\(\notationII \isacommand{typedef}\)-generated function \(\notationII \definedterm{expr}\) is undefined on terms with
dangling indices.  These premises could be eliminated by defining a
more tightly-specified version of \(\notationII \definedterm{expr}\), satisfying the same
\(\notationII \isacommand{typedef}\)-generated axioms while preserving the structure of its
argument except for any dangling indices.  This was done in the previous
version of Hybrid~\cite{momigliano/martin/felty:2008}
(with the help of an auxiliary function called \(\notationII \definedterm{trim}\)).  However, with a
more systematic treatment of \(\notationII \definedterm{level}\) and some additional lemmas for it,
this was found to be unnecessary.

All versions of Hybrid follow a general pattern of making definitions and
proving lemmas first for arbitrary levels, and then deriving the desired
results for proper terms as corollaries.  In the present version, arbitrary
levels are handled by recursion and induction over de~Bruijn syntax, using
the type \(\notationII \isaholtype{dB}\) and the predicate \(\notationII \definedterm{level}\), while the results for proper
terms are stated at type \(\notationII \isaholtype{expr}\).

\section{Definition of \ldquo abstr\rdquo\ and \ldquo LAM\rdquo}
\label{sec:hybrid-defn-abstr-LAM}

We now turn to the task of defining \(\notationII \definedterm{abstr}\) and \(\notationII \definedterm{LAM}\).
The main ideas are from \cite{ambler/crole/momigliano:2002}, but the
details of the definitions and proofs are original.  There are some
improvements over the original version of Hybrid, which will be described
in this section and \abbrevref{sec:hybrid-defn-LAM-inject}.

Since we will be defining \(\notationII \definedterm{abstr}\) and \(\notationII \definedterm{LAM}\) in terms of de~Bruijn
syntax, the definition of syntactic functions from
\abbrevref{sec:hybrid-overview} is not directly usable here:
we need an analogous definition using de~Bruijn syntax in place of \(\notationII \definedterm{LAM}\).

For recursion, we must work with \(\notationII \isaholtype{dB}\)-valued functions (arbitrary levels)
rather than \(\notationII \isaholtype{expr}\)-valued functions.  However, the argument type need not
also be \(\notationII \isaholtype{dB}\), and in fact it will be more convenient to work with functions
of type \(\notationII  (\isaholtype{expr} \isatypearrow \isaholtype{dB}) \).  This simplifies the treatment of \(\notationII \definedterm{level}\)
by avoiding negative occurrences of the type \(\notationII \isaholtype{dB}\).

Thus we define the \defterm{syntactic \(\notationIII \isaholtype{dB}\)-terms}, as a subset of
Isabelle\slash HOL terms of type \(\notationIII \isaholtype{dB}\), using variables of type \(\notationIII \isaholtype{expr}\)
converted via \(\notationIII \definedterm{dB}\):
  \[\notationIII  s ::= \definedterm{dB}\ x \mid \definedterm{CON'}\ a \mid \definedterm{VAR'}\ n \mid s_1 \hybridappO s_2 \mid %
                    \definedterm{ERR'} \mid \definedterm{BND'}\ i \mid \definedterm{ABS'}\ s \]
where \(\notationIII s\) (with possible subscripts) stands for a syntactic \(\notationIII \isaholtype{dB}\)-term,
\(\notationIII x\) for a variable of type \(\notationIII \isaholtype{expr}\), \(\notationIII a\) for a constant of
type \(\notationIII \isaholtype{con}\), and \(\notationIII n\) and \(\notationIII i\) for natural-number constants.
We define the \defterm{syntactic \(\notationIII \isaholtype{dB}\)-functions} as the functions of type
\(\notationIII  (\isaholtype{expr} \isatypearrow \isaholtype{dB}) \) of the form \(\notationIII  (\isalambda  \localvar{x} \ldotp s) \), where \(\notationIII s\) is
a syntactic \(\notationIII \isaholtype{dB}\)-term with (at most) one free variable \(\notationIII \localvar{x}\).
Such functions mix de~Bruijn indices (\(\notationIII \definedterm{BND'}\)) with HOAS (using the
Isabelle\slash HOL bound variable \(\notationIII \localvar{x}\) to represent an object-language
variable).

We define a predicate \(\notationII \definedterm{Abstr}\) to recognize the syntactic
\(\notationIII \isaholtype{dB}\)-functions, which formally defines the so-far only
informally identified set.  We also define an auxiliary predicate
\(\notationII \definedterm{ordinary}\) needed in the definition of \(\notationII \definedterm{Abstr}\):

\begin{Defn}
\label{defn:hybrid-ordinary} \formal
\(\notationII  \negquad \definedterm{ordinary} \isaoftype (\isatypevar{b} \isatypearrow \isatypevar{a}\ \isaholtype{dB}) \isatypearrow \isaholtype{bool} \)\par\nopagebreak
\(\notationII   \definedterm{ordinary}\ \localvar{X} { \; \isaequiv \quad } %
         ( \isaholexists\ \localvar{a}. \localvar{X} = (\isalambda\ \localvar{x}. \definedterm{CON'}\ \localvar{a})) %
    \isaholdisj ( \isaholexists\ \localvar{n}. \localvar{X} = (\isalambda\ \localvar{x}. \definedterm{VAR'}\ \localvar{n})) \isaholdisj {} \)\par\nopagebreak
\(\notationII  \phantom{ \definedterm{ordinary}\ \localvar{X} { \; \isaequiv \quad} } %
        ( \isaholexists\ \localvar{S}\ \localvar{T}. \localvar{X} = (\isalambda\ \localvar{x}. \localvar{S}\ \localvar{x} \hybridappO \localvar{T}\ \localvar{x})) %
   \isaholdisj (\localvar{X} = (\isalambda\ \localvar{x}. \definedterm{ERR'})) \isaholdisj {} \)\par\nopagebreak
\(\notationII  \phantom{ \definedterm{ordinary}\ \localvar{X} { \; \isaequiv \quad } } %
         ( \isaholexists\ \localvar{j}. \localvar{X} = (\isalambda\ \localvar{x}. \definedterm{BND'}\ \localvar{j})) %
    \isaholdisj ( \isaholexists\ \localvar{S}. \localvar{X} = (\isalambda\ \localvar{x}. \definedterm{ABS'}\ (\localvar{S}\ \localvar{x}))) \)
\end{Defn}

\begin{Defn}
\label{defn:hybrid-Abstr} \formal
\(\notationII  \negquad \isacommand{function}\ \definedterm{Abstr} \isaoftype (\isatypevar{a}\ \isaholtype{expr} \isatypearrow \isatypevar{a}\ \isaholtype{dB}) \isatypearrow \isaholtype{bool} \)\par\nopagebreak
\(\notationII    \definedterm{Abstr}\ (\isalambda\ \localvar{x}. s) = \definedterm{True} \)
      where \(\notationII s\) is\/ \(\notationII  (\definedterm{CON'}\ \localvar{a}) \), \(\notationII  (\definedterm{VAR'}\ \localvar{n}) \), \(\notationII  \definedterm{ERR'} \),
      or\/ \(\notationII  (\definedterm{BND'}\ \localvar{i}) \)\par\nopagebreak
\(\notationII    \definedterm{Abstr}\ (\isalambda\ \localvar{x}. \localvar{S}\ \localvar{x} \hybridappO \localvar{T}\ \localvar{x}) = (\definedterm{Abstr}\ \localvar{S} \isaholconj \definedterm{Abstr}\ \localvar{T}) \)\par\nopagebreak[3]
\(\notationII    \definedterm{Abstr}\ (\isalambda\ \localvar{x}. \definedterm{ABS'}\ (\localvar{S}\ \localvar{x})) = \definedterm{Abstr}\ \localvar{S} \)\par\nopagebreak
\(\notationII    \neg\ \definedterm{ordinary}\ \localvar{S} \isaimpl \definedterm{Abstr}\ \localvar{S} = (\localvar{S} = \definedterm{dB}) \)\par\smallskip\pagebreak[0]
\end{Defn}

Syntactically, the defining equations for \(\notationII \definedterm{Abstr}\) have the form of
recursion on the \emph{body} of a \hbox{$\lambda$-abstraction}.
Mathematically, they define \(\notationII  (\definedterm{Abstr}\ \localvar{S}) \) by recursion on the
\emph{common structure} of all the values of the function \(\notationII \localvar{S}\), i.e.,
on the common structure (if any) of \(\notationII  (\localvar{S}\ \localvar{x}) \) for all \(\notationII  \localvar{x} \isaoftype \isaholtype{expr} \).
The predicate \(\notationII \definedterm{ordinary}\) recognizes those functions that match one of
the first three equations, so that the condition \(\notationII  ( \neg\ \definedterm{ordinary}\ \localvar{S}) \)
on the last equation may be read as ``otherwise''; that equation corresponds
to the variable case for syntactic \(\notationII \isaholtype{dB}\)-terms as defined above.

This definition is formalized with the help of Isabelle\slash HOL's
\(\notationII \isacommand{function}\) command.  It demands proofs of pattern completeness,
compatibility, and termination (not shown), and then in addition to
defining \(\notationII \definedterm{Abstr}\) and proving its defining equations, it automatically
generates structural induction and case-distinction rules for the type
\(\notationII  (\isaholtype{expr} \isatypearrow \isaholtype{dB}) \) corresponding to the pattern of recursion used
in the definition; these are called \(\notationII  \isatheorem{Abstr\ldotsep induct} \) and
\(\notationII  \isatheorem{Abstr\ldotsep cases} \) respectively, and will be referred to
later.

We may now define the predicate \(\notationII \definedterm{abstr}\) in terms of \(\notationII \definedterm{Abstr}\) by
using post-\break composition with \(\notationII \definedterm{dB}\) to convert its function argument
from the type \(\notationII  (\isaholtype{expr} \isatypearrow \isaholtype{expr}) \) to \(\notationII { (\isaholtype{expr} \isatypearrow \isaholtype{dB}) }\).

\begin{Defn}
\label{defn:hybrid-abstr} \formal
\(\notationII  \negquad \definedterm{abstr} \isaoftype (\isatypevar{a}\ \isaholtype{expr} \isatypearrow \isatypevar{a}\ \isaholtype{expr}) \isatypearrow \isaholtype{bool} \)\par\nopagebreak
\(\notationII  \definedterm{abstr}\ \localvar{S} \isaequiv \definedterm{Abstr}\ (\definedterm{dB} \circ \localvar{S}) \)
\end{Defn}

Note that unlike the situation in \cite{ambler/crole/momigliano:2002},
the definition of \(\notationII \definedterm{Abstr}\) does not need to impose a constraint on the
argument of \(\notationII \definedterm{BND'}\), because in the case of \(\notationII  (\definedterm{abstr}\ \localvar{S}) \) dangling
indices are excluded by the type of the function \(\notationII  \localvar{S} \isaoftype (\isaholtype{expr} \isatypearrow \isaholtype{expr}) \).

\begin{Lem}
\label{lem:hybrid-Abstr-const} \formal
\(\notationII  \negquad \isatheorem{Abstr\_const} 
   \colon \definedterm{Abstr}\ (\isalambda\ \localvar{x}. \localvar{s}) \)
\end{Lem}

The lemma \(\notationII \isatheorem{Abstr\_const}\) shows that any constant function of type
\(\notationII  (\isaholtype{expr} \isatypearrow \isaholtype{dB}) \) satisfies \(\notationII \definedterm{Abstr}\).  It is used to prove a similar
property for \(\notationII \definedterm{abstr}\), and will later be used directly as well.
It is proved by induction on \(\notationII \localvar{s}\) using \abbrevref{defn:hybrid-Abstr}
(\(\notationII \definedterm{Abstr}\)).

\begin{Lem}
\label{lem:hybrid-abstr-simps-part1} \formal 
\(\notationII  \negquad \isatheorem{abstr\_id}\colon  \definedterm{abstr}\ (\isalambda\ \localvar{x}. \localvar{x}) \)\par\nopagebreak[3]
\(\notationII  \negquad \isatheorem{abstr\_const}\colon \definedterm{abstr}\ (\isalambda\ \localvar{x}. \localvar{s}) \)\par\nopagebreak[3]
\(\notationII  \negquad \isatheorem{abstr\_APP}\colon %
     \definedterm{abstr}\ (\isalambda\ \localvar{x}. \localvar{S}\ \localvar{x} \hybridapp \localvar{T}\ \localvar{x}) \isaholiff (\definedterm{abstr}\ \localvar{S} \isaholconj \definedterm{abstr}\ \localvar{T}) \)
\end{Lem}

The lemma \(\notationII \isatheorem{abstr\_const}\) is a corollary of \(\notationII \isatheorem{Abstr\_const}\),
while the other two lemmas are proved directly, using
Definitions \ref{defn:hybrid-abstr} (\(\notationII \definedterm{abstr}\))
and \ref{defn:hybrid-Abstr} (\(\notationII \definedterm{Abstr}\)).

These lemmas allow \(\notationII \definedterm{abstr}\) conditions for syntactic functions to be
proved compositionally without unfolding the definition, except when
the body of the function contains a \(\notationII \definedterm{LAM}\) subterm that involves the
function argument (so that it is not just a constant).  In that case,
previous versions of Hybrid required unfolding the definitions of \(\notationII \definedterm{abstr}\)
and \(\notationII \definedterm{LAM}\) to convert HOAS to de~Bruijn syntax.  The present work improves
on that situation by providing a compositional rule also for the \(\notationII \definedterm{LAM}\)
case (\abbrevref{lem:hybrid-abstr-LAM} in
\abbrevref{sec:hybrid-abstr2-and-abstr-LAM}).

The lemma \(\notationII \isatheorem{abstr\_const}\) will be important for Hybrid terms with nested
\(\notationII \definedterm{LAM}\) operators, to show that the argument of an inner \(\notationII \definedterm{LAM}\) satisfies
\(\notationII \definedterm{abstr}\) when its body contains a bound variable from an outer \(\notationII \definedterm{LAM}\);
such a bound variable is a placeholder for an arbitrary term of type
\(\notationII \isaholtype{expr}\), which is exactly the role of \(\notationII \localvar{s}\) in \(\notationII \isatheorem{abstr\_const}\).

We now define the function \(\notationII \definedterm{LAM}\), using the same form of recursion
that was used in the definition of \(\notationII \definedterm{abstr}\).

\begin{Defn}
\label{defn:hybrid-LAM} \formal
\(\notationII  \negquad \definedterm{LAM} \isaoftype (\isatypevar{a}\ \isaholtype{expr} \isatypearrow \isatypevar{a}\ \isaholtype{expr}) \isatypearrow \isatypevar{a}\ \isaholtype{expr} \)\par\nopagebreak
\(\notationII  \definedterm{LAM}\ \localvar{S} \isaequiv \definedterm{expr}\ (\definedterm{Lambda}\ (\definedterm{dB} \circ \localvar{S})) \)\par\nopagebreak[3]
\(\notationII  \negquad \definedterm{Lambda} \isaoftype (\isatypevar{a}\ \isaholtype{expr} \isatypearrow \isatypevar{a}\ \definedterm{dB}) \isatypearrow \isatypevar{a}\ \isaholtype{dB} \)\par\nopagebreak
\(\notationII  \definedterm{Lambda}\ \localvar{S} \isaequiv %
     \definedterm{if}\ (\definedterm{Abstr}\ \localvar{S})\ \definedterm{then}\ (\definedterm{ABS'}\ (\definedterm{Lbind}\ 0\ \localvar{S}))\ \definedterm{else}\ \definedterm{ERR'} \)
\end{Defn}

The function \(\notationII \definedterm{LAM}\), like \(\notationII \definedterm{abstr}\), first composes  \(\notationII \definedterm{dB}\) with the
given function.  It then applies the auxiliary function \(\notationII \definedterm{Lambda}\) and
converts the resulting term from type \(\notationII \isaholtype{dB}\) to type \(\notationII \isaholtype{expr}\).

The function \(\notationII \definedterm{Lambda}\) first checks if its argument satisfies \(\notationII \definedterm{Abstr}\),
and produces \(\notationII \definedterm{ERR'}\) if not.  (This is equivalent to checking if the
argument of \(\notationII \definedterm{LAM}\) satisfies \(\notationII \definedterm{abstr}\).)  The original version of
Hybrid~\cite{ambler/crole/momigliano:2002} did not do this check (and did
not have the constant \(\notationII \definedterm{ERR'}\)), making it impossible to determine from
\(\notationII  (\definedterm{LAM}\ \localvar{S}) \) whether \(\notationII \localvar{S}\) was a syntactic function or not.  We include
these features to support the stronger injectivity property for \(\notationII \definedterm{LAM}\)
proved in \abbrevref{sec:hybrid-defn-LAM-inject}.

If its argument does satisfy \(\notationII \definedterm{Abstr}\), then \(\notationII \definedterm{Lambda}\) applies another
auxiliary function \(\notationII \definedterm{Lbind}\), defined by recursion, to convert HOAS to
de~Bruijn syntax; i.e., to convert the variable represented by the
function argument into a dangling de~Bruijn index.  It then applies a
new \(\notationII \definedterm{ABS'}\) node to bind the variable and obtain a proper de~Bruijn~term.

\begin{Defn}
\label{defn:hybrid-Lbind} \formal
\(\notationII  \negquad \isacommand{function} %
  \ \definedterm{Lbind} \isaoftype \isabiglbrack \isaholtype{bnd}, (\isatypevar{a}\ \isaholtype{expr} \isatypearrow \isatypevar{a}\ \isaholtype{dB}) \isabigrbrack \isatypearrow \isatypevar{a}\ \isaholtype{dB} \)\par\nopagebreak
\(\notationII    \definedterm{Lbind}\ \localvar{i}\ (\isalambda\ \localvar{x}. s) = s \)
      where \(\notationII s\) is\/ \(\notationII  (\definedterm{CON'}\ \localvar{a}) \), \(\notationII  (\definedterm{VAR'}\ \localvar{n}) \), \(\notationII  \definedterm{ERR'} \),
      or\/ \(\notationII  (\definedterm{BND'}\ \localvar{j}) \)\par\nopagebreak
\(\notationII    \definedterm{Lbind}\ \localvar{i}\ (\isalambda\ \localvar{x}. \localvar{S}\ \localvar{x} \hybridappO \localvar{T}\ \localvar{x}) = \definedterm{Lbind}\ \localvar{i}\ \localvar{S} \hybridappO \definedterm{Lbind}\ \localvar{i}\ \localvar{T} \)\par\nopagebreak[3]
\(\notationII    \definedterm{Lbind}\ \localvar{i}\ (\isalambda\ \localvar{x}. \definedterm{ABS'}\ (\localvar{S}\ \localvar{x})) = \definedterm{ABS'}\ (\definedterm{Lbind}\ (\localvar{i} + 1)\ \localvar{S}) \)\par\nopagebreak
\(\notationII    \neg\ \definedterm{ordinary}\ \localvar{S} \isaimpl \definedterm{Lbind}\ \localvar{i}\ \localvar{S} = \definedterm{BND'}\ \localvar{i} \) 
\end{Defn}

The auxiliary function \(\notationII \definedterm{Lbind}\) extracts the common structure of the
values of its function argument, replacing indecomposable uses of the bound
variable (i.e., functions that do not match any of the first three equations)
with \(\notationII  (\definedterm{BND'}\ \localvar{i}) \).  This is a dangling de~Bruijn index, and \(\notationII \localvar{i}\) is
incremented each time the recursion passes an \(\notationII \definedterm{ABS'}\) node so that all
such instances of \(\notationII \definedterm{BND'}\) will refer to the \(\notationII \definedterm{ABS'}\) node added by
\(\notationII \definedterm{Lambda}\).  The \(\notationII \definedterm{Abstr}\) condition checked in the definition of
\(\notationII \definedterm{Lambda}\) ensures that the last equation will be applied only when
\(\notationII  \localvar{S} = (\isalambda\ \localvar{x}. \definedterm{dB}\ \localvar{x}) \).

\begin{Lem}
\label{lem:hybrid-Lbind-const} \formal
\(\notationII  \negquad \isatheorem{Lbind\_const} 
   \colon \definedterm{Lbind}\ \localvar{i}\ (\isalambda\ \localvar{x}. \localvar{s}) = \localvar{s} \)
\end{Lem}

The lemma \(\notationII \isatheorem{Lbind\_const}\) shows that applying \(\notationII  (\definedterm{Lbind}\ \localvar{i}) \) to
a constant function of type \(\notationII  (\isaholtype{expr} \isatypearrow \isaholtype{dB}) \) gives the constant value
of that function.  It is proved by induction on \(\notationII \localvar{s}\).
This lemma will be important for Hybrid terms with nested \(\notationII \definedterm{LAM}\)
operators, to allow the argument of an outer \(\notationII \definedterm{LAM}\) to satisfy \(\notationII \definedterm{abstr}\)
when its bound variable occurs in the scope of an inner \(\notationII \definedterm{LAM}\).

\begin{Lem}
\label{lem:hybrid-dB-LAM} \formal
\(\notationII  \negquad \isatheorem{dB\_LAM} 
   \colon \definedterm{dB}\ (\definedterm{LAM}\ \localvar{S}) = \definedterm{if}\ (\definedterm{abstr}\ \localvar{S})\ \definedterm{then}\ (\definedterm{ABS'}\ (\definedterm{Lbind}\ 0\ (\definedterm{dB} \circ \localvar{S})))\ \definedterm{else}\ \definedterm{ERR'} \)\par\nopagebreak
\(\notationII  \negquad \isatheorem{abstr\_dB\_LAM} 
   \colon \definedterm{abstr}\ \localvar{S} \isaimpl \definedterm{dB}\ (\definedterm{LAM}\ \localvar{S}) = \definedterm{ABS'}\ (\definedterm{Lbind}\ 0\ (\definedterm{dB} \circ \localvar{S})) \)
\end{Lem}

The lemma \(\notationII \isatheorem{dB\_LAM}\) combines unfolding of \abbrevref{defn:hybrid-LAM}
(\(\notationII \definedterm{LAM}\) and \(\notationII \definedterm{Lambda}\)) with cancellation of the functions \(\notationII \definedterm{dB}\) and
\(\notationII \definedterm{expr}\), using the fact that both \(\notationII \definedterm{ERR'}\) and
\(\notationII  (\definedterm{ABS'}\ (\definedterm{Lbind}\ 0\ (\definedterm{dB} \circ \localvar{S}))) \) are proper.  (Dangling indices are
excluded from \(\notationII  \localvar{S} \isaoftype (\isaholtype{expr} \isatypearrow \isaholtype{expr}) \) by its type, and the one introduced
by \(\notationII \definedterm{Lbind}\) is bound by the enclosing~\(\notationII \definedterm{ABS'}\).)
The lemma \(\notationII \isatheorem{abstr\_dB\_LAM}\) is a weaker version intended as a conditional
rewrite rule for Isabelle's simplifier, to do the unfolding only if the
\(\notationII \definedterm{abstr}\) condition simplifies to \(\notationII \definedterm{True}\).

With the definitions above, Hybrid terms using \(\notationII \definedterm{LAM}\)\ (i.e., closed
syntactic terms) are provably equal to the corresponding de~Bruijn syntax
representations, converted to the type \(\notationII \isaholtype{expr}\) using the function
\(\notationII \definedterm{expr}\).  (This is much the same situation as in
\cite{ambler/crole/momigliano:2002}, except for the type conversion
which was not necessary there.)  Thus, starting from two \emph{distinct}
representations for free variables, we have established two \emph{ambiguous}
representations for bound variables, in the sense that any given element
of \(\notationII \isaholtype{expr}\) may be viewed as having either form.
In the following sections, we will state results using
the HOAS representation~(\(\notationII \definedterm{LAM}\)) but use the de~Bruijn syntax
representation~(\(\notationII \definedterm{ABS'}\)/\(\notationII \definedterm{BND'}\)) in proofs by induction, aiming to
characterize the former representation so that it stands on its own.

All versions of Hybrid have used essentially the same form of recursion to
define \(\notationII \definedterm{abstr}\) and \(\notationII \definedterm{LAM}\), and the corresponding form of induction
to prove their properties.  However, the means of formalizing it have varied
greatly.  The original version \cite{ambler/crole/momigliano:2002} used
inductively-defined predicates and induction on those predicates; the
following version \cite{momigliano/martin/felty:2008}
used primitive recursion and induction on an auxiliary datatype \(\notationII \isaholtype{dB\_fn}\);
while the present version avoids many of the complications of the previous
approaches with the help of the \(\notationII \isacommand{function}\) command.

A predicate called \(\notationII \definedterm{ordinary}\) has also been present in all versions of
Hybrid, though it originally included the variable case as well.
Removing this case allowed \(\notationII \definedterm{ordinary}\) to be generalized to \(\notationII \isaholtype{dB}\)-valued
functions on any type; this will allow us to reuse it for binary functions in
\abbrevref{sec:hybrid-abstr2-and-abstr-LAM}.  (It is also reused for $n$-ary
functions in \cite[Sect.~3.3]{martin:2010a}.)

\section{Injectivity of \ldquo LAM\rdquo}
\label{sec:hybrid-defn-LAM-inject}

As stated in \abbrevref{sec:hybrid-overview}, Hybrid proves injectivity of
\(\notationII \definedterm{LAM}\) restricted to functions of type \(\notationII  (\isaholtype{expr} \isatypearrow \isaholtype{expr}) \) satisfying
\(\notationII \definedterm{abstr}\).  Improving on \cite{ambler/crole/momigliano:2002}, this property
is strengthened by requiring only one \(\notationII \definedterm{abstr}\) premise, using the fact
that \(\notationII \definedterm{LAM}\) maps functions not satisfying \(\notationII \definedterm{abstr}\) to a recognizable
placeholder term \(\notationII \definedterm{ERR}\).

We begin with an injectivity result for arbitrary de~Bruijn levels.
To state this result concisely, we first define an abbreviation \(\notationII \definedterm{Level}\)
for pointwise application of \(\notationII \definedterm{level}\) to a function:

\begin{Defn}
\label{defn:hybrid-Level} \formal
\(\notationII  \negquad \isacommand{abbreviation} %
  \ \definedterm{Level} \isaoftype \isabiglbrack \isaholtype{bnd}, (\isatypevar{b} \isatypearrow \isatypevar{a}\ \isaholtype{dB}) \isabigrbrack \isatypearrow \isaholtype{bool} \)\par\nopagebreak[3]
\(\notationII    \definedterm{Level}\ \localvar{i}\ \localvar{S} \isaequiv  \isaholuniv\ \localvar{x}. \definedterm{level}\ \localvar{i}\ (\localvar{S}\ \localvar{x}) \)
\end{Defn}

\begin{Lem}
\label{lem:hybrid-abstr-lbind-inject} \formal
\(\notationII  \negquad \isatheorem{Abstr\_Lbind\_inject}\colon \)\par\nopagebreak 
\(\notationII  \isabigldblbrack \mkern-1mu \definedterm{Abstr}\ \localvar{S}; \definedterm{Abstr}\ \localvar{T}; %
      \definedterm{Level}\ \localvar{i}\ \localvar{S}; \definedterm{Level}\ \localvar{i}\ \localvar{T} \mkern-1mu \isabigrdblbrack \mkern-1mu \isaimpl \mkern-1mu %
     (\definedterm{Lbind}\ \localvar{i}\ \localvar{S} \mkern-1mu = \mkern-1mu \definedterm{Lbind}\ \localvar{i}\ \localvar{T}) %
     \mkern-1mu \isaholiff \mkern-1mu (\localvar{S} \mkern-1mu = \mkern-1mu \localvar{T}) \)
\end{Lem}

This lemma is proved by a straightforward  induction on
\(\notationII  \localvar{S} \isaoftype (\isaholtype{expr} \isatypearrow \isaholtype{dB}) \) using \(\notationII \isatheorem{Abstr\ldotsep induct}\)
(from \abbrevref{defn:hybrid-Abstr}).

\begin{Theo}[Injectivity of \(\notationII \definedterm{LAM}\)]
\label{theo:hybrid-LAM-inject} \formal
\(\notationII  \isabigldblbrack \definedterm{LAM}\ \localvar{S} = \definedterm{LAM}\ \localvar{T}; \definedterm{abstr}\ \localvar{S} \isaholdisj \definedterm{abstr}\ \localvar{T} \isabigrdblbrack \isaimpl \localvar{S} = \localvar{T} \)
\end{Theo}

\begin{Proof}
If one of \(\notationII \localvar{S}\) and \(\notationII \localvar{T}\) satisfies \(\notationII \definedterm{abstr}\) and the other does not,
then by \abbrevref{lem:hybrid-dB-LAM} (\(\notationII \isatheorem{dB\_LAM}\)), one of the terms
\(\notationII  (\definedterm{dB}\ (\definedterm{LAM}\ \localvar{S})) \) and \(\notationII  (\definedterm{dB}\ (\definedterm{LAM}\ \localvar{T})) \) is of the form
\(\notationII  (\definedterm{ABS'}\ \localvar{t}) \) for some \(\notationII  \localvar{t} \isaoftype \isaholtype{dB} \), while the other is \(\notationII \definedterm{ERR'}\).
But these terms cannot be equal, which contradicts the premise
\(\notationII  \definedterm{LAM}\ \localvar{S} = \definedterm{LAM}\ \localvar{T} \).  Thus the original assumption must be false,
and we must have both \(\notationII  (\definedterm{abstr}\ \localvar{S}) \) and \(\notationII  (\definedterm{abstr}\ \localvar{T}) \).

We apply \(\notationII \definedterm{dB}\) to both sides of the equality \(\notationII  \definedterm{LAM}\ \localvar{S} = \definedterm{LAM}\ \localvar{T} \)
and simplify using \(\notationII \isatheorem{abstr\_dB\_LAM}\)\ (\abbrevref{lem:hybrid-dB-LAM})
to obtain
  \[\notationII  \definedterm{ABS'}\ (\definedterm{Lbind}\ 0\ (\definedterm{dB} \circ \localvar{S})) = \definedterm{ABS'}\ (\definedterm{Lbind}\ 0\ (\definedterm{dB} \circ \localvar{T})) \ldotp \]
\(\notationII \definedterm{ABS'}\) is a datatype constructor and thus injective, so we may cancel it:
  \[\notationII  \definedterm{Lbind}\ 0\ (\definedterm{dB} \circ \localvar{S}) = \definedterm{Lbind}\ 0\ (\definedterm{dB} \circ \localvar{T}) \ldotp \]
We have \(\notationII  (\definedterm{Abstr}\ (\definedterm{dB} \circ \localvar{S})) \) and \(\notationII  (\definedterm{Abstr}\ (\definedterm{dB} \circ \localvar{T})) \) by
unfolding \abbrevref{defn:hybrid-abstr} (\(\notationII \definedterm{abstr}\)), and we also have
\(\notationII  (\definedterm{Level}\ 0\ (\definedterm{dB} \circ \localvar{S})) \) and \(\notationII  (\definedterm{Level}\ 0\ (\definedterm{dB} \circ \localvar{T})) \) since terms
converted from type \(\notationII \isaholtype{expr}\) are proper by \autoref{defn:hybrid-expr-type}.
Thus we may apply the preceding lemma (\(\notationII \isatheorem{Abstr\_Lbind\_inject}\)) to deduce
\(\notationII  \definedterm{dB} \circ \localvar{S} = \definedterm{dB} \circ \localvar{T} \).  Since \(\notationII \definedterm{dB}\) is injective, it can be canceled
to obtain \(\notationII  \localvar{S} = \localvar{T} \), as was to be proven.
\end{Proof}

Note that \(\notationII  (\definedterm{Lbind}\ 0) \) is only injective on functions from \(\notationII \isaholtype{expr}\)
to \(\notationII \isaholtype{dB}\) whose values are proper terms, i.e., those that factor through
\(\notationII \definedterm{dB}\), because any pre-existing dangling indices at level 1 would be
indistinguishable from those resulting from conversion of the HOAS variable.
For example,
  \[\notationII  \definedterm{Lbind}\ 0\ (\isalambda\ \localvar{x}. \definedterm{dB}\ \localvar{x}) = \definedterm{BND'}\ 0 = \definedterm{Lbind}\ 0\ (\isalambda\ \localvar{x}. \definedterm{BND'}\ 0)
     \ldotp \]
Thus, without the \(\notationII \isacommand{typedef}\) limiting \(\notationII \isaholtype{expr}\) to proper terms,
we would not be able to avoid conditions on both \(\notationII \localvar{S}\) and \(\notationII \localvar{T}\);
at best, we could replace one \(\notationII \definedterm{abstr}\) condition
with something like \(\notationII  ( \isaholuniv\ \localvar{x}. \definedterm{proper}\ \localvar{x} \isaholimpl \definedterm{proper}\ (\localvar{T}\ \localvar{x})) \).

The advantage of an injectivity property that can work with a condition on
only one of \(\notationII \localvar{S}\) and \(\notationII \localvar{T}\) is that it simplifies the elimination rules
for inductively-defined predicates on Hybrid terms, such as the formalization
of evaluation for Mini-ML with references in \cite[Sect.~5.3]{martin:2010a}.
As a result, \(\notationII \definedterm{abstr}\) conditions are more often available where they are
needed, without having to add them as premises.

Distinctness of \(\notationII \definedterm{LAM}\) from the first-order operators of
\autoref{defn:hybrid-fo-ops} follows straightforwardly from
\autoref{defn:hybrid-LAM}, except that \(\notationII  (\definedterm{LAM}\ \localvar{F}) \) is distinct from
\(\notationII \definedterm{ERR}\) only under the premise \(\notationII  (\definedterm{abstr}\ \localvar{F}) \).

\section{Characterizing \ldquo abstr\rdquo}
\label{sec:hybrid-abstr2-and-abstr-LAM}

In \abbrevref{sec:hybrid-defn-abstr-LAM}, an incomplete set of simplification
rules for \(\notationII \definedterm{abstr}\) was provided as \abbrevref{lem:hybrid-abstr-simps-part1}.
The missing case is \(\notationII  (\definedterm{abstr}\ (\isalambda\ \localvar{x}. \definedterm{LAM}\ \localvar{y}. \localvar{W}\ \localvar{x}\ \localvar{y})) \).

Both previous versions of Hybrid
\cite{ambler/crole/momigliano:2002,momigliano/martin/felty:2008}
relied on conversion from HOAS to de~Bruijn syntax to handle this case.
That is sufficient for proving that particular syntactic functions satisfy
\(\notationII \definedterm{abstr}\),
but it is less useful for partially-specified functions as found in
inductive proofs.

We could obtain a compositional introduction rule for this case by defining
a predicate
  \(\notationII  \definedterm{biAbstr} \isaoftype \bigl( {[} \isaholtype{expr}, \isaholtype{expr} {]} \isatypearrow \isaholtype{expr} \bigr) \isatypearrow \isaholtype{bool} \)
generalizing \(\notationII \definedterm{abstr}\), and proving
  \[\notationII  \definedterm{biAbstr}\ \localvar{W} \isaimpl \definedterm{abstr}\ (\isalambda\ \localvar{x}. \definedterm{LAM}\ \localvar{y}. \localvar{W}\ \localvar{x}\ \localvar{y}) \ldotp \]
This was done by Momigliano et al.\ \cite{momigliano/ambler/crole:2002};
their formal theory \(\notationII \definedterm{BiAbstr}\) is available online~%
\cite{felty/momigliano:2010}.
However, the \(\notationII \definedterm{LAM}\) case arises again for \(\notationII \definedterm{biAbstr}\), and for any
higher-arity generalization.  There are several ways to address this:

\begin{itemize}
\item Use Isabelle\slash HOL's axiomatic type classes to define a polymorphic
  predicate generalizing \(\notationII \definedterm{abstr}\) to curried functions of arbitrary arity.
  This looks like a promising approach, but it remains as future work.
\item Find a single type that can represent functions of arbitrary arity,
  and generalize Hybrid's constructs to that type.  (Some experimental work
  has been done in that direction \cite[Sect.~3.3]{martin:2010a}.)  Such a
  type is also useful as a representation of open terms for induction.
\item Prove a result that reduces \(\notationII \definedterm{biAbstr}\) to \(\notationII \definedterm{abstr}\).  This
  seems to be the most direct solution, and it is the approach we take
  in the present work.
\end{itemize}

In this section, we will represent functions of two arguments using pairs,
rather than in the usual curried form, so that we may reuse
\abbrevref{defn:hybrid-ordinary} (\(\notationII \definedterm{ordinary}\)) and some technical lemmas
(left unstated as they are mathematically trivial), all of which refer to
the polymorphic type \(\notationII  (\isatypevar{b} \isatypearrow \isaholtype{dB}) \).

\begin{Defn} \formal
\(\notationII  \negquad \definedterm{abstr\_2} \isaoftype (\isatypevar{a}\ \isaholtype{expr} \isapairtype \isatypevar{a}\ \isaholtype{expr} \isatypearrow \isatypevar{a}\ \isaholtype{expr}) \isatypearrow \isaholtype{bool} \)\par\nopagebreak
\(\notationII  \definedterm{abstr\_2}\ \localvar{S} \isaequiv \definedterm{Abstr\_2}\ (\definedterm{dB} \circ \localvar{S}) \)
\end{Defn}

The predicate \(\notationII \definedterm{abstr\_2}\) generalizes \(\notationII \definedterm{abstr}\) to functions on the
Cartesian product type \(\notationII  (\isaholtype{expr} \isapairtype \isaholtype{expr}) \); it corresponds to
\(\notationII \definedterm{biAbstr}\) \cite{momigliano/ambler/crole:2002}.  It is defined in
the same way as \(\notationII \definedterm{abstr}\), composing \(\notationII \definedterm{dB}\) with its argument and
then applying a recursively-defined auxiliary predicate \(\notationII \definedterm{Abstr\_2}\).

\begin{Defn} \formal
\(\notationII  \negquad \isacommand{function} %
  \ \definedterm{Abstr\_2} \isaoftype (\isatypevar{a}\ \isaholtype{expr} \isapairtype \isatypevar{a}\ \isaholtype{expr} \isatypearrow \isatypevar{a}\ \isaholtype{dB}) \isatypearrow \isaholtype{bool} \)\par\nopagebreak
\(\notationII    \definedterm{Abstr\_2}\ (\isalambda\ \localvar{p}. s) = \definedterm{True} \)
      where \(\notationII s\) is\/ \(\notationII  (\definedterm{CON'}\ \localvar{a}) \), \(\notationII  (\definedterm{VAR'}\ \localvar{n}) \), \(\notationII  \definedterm{ERR'} \),
      or\/ \(\notationII  (\definedterm{BND'}\ \localvar{i}) \)\par\nopagebreak
\(\notationII    \definedterm{Abstr\_2}\ (\isalambda\ \localvar{p}. \localvar{S}\ \localvar{p} \hybridapp \localvar{T}\ \localvar{p}) = (\definedterm{Abstr\_2}\ \localvar{S} \isaholconj \definedterm{Abstr\_2}\ \localvar{T}) \)\par\nopagebreak[3]
\(\notationII    \definedterm{Abstr\_2}\ (\isalambda\ \localvar{p}. \definedterm{ABS'}\ (\localvar{S}\ \localvar{p})) = \definedterm{Abstr\_2}\ \localvar{S} \)\par\nopagebreak
\(\notationII    \neg\ \definedterm{ordinary}\ \localvar{S} \isaimpl \definedterm{Abstr\_2}\ \localvar{S} = (\localvar{S} = \definedterm{dB} \circ \definedterm{fst} \isaholdisj \localvar{S} = \definedterm{dB} \circ \definedterm{snd}) \)
\end{Defn}

The predicate \(\notationII \definedterm{Abstr\_2}\) is similar to \(\notationII \definedterm{Abstr}\), except that it has
\emph{two} variable cases: \(\notationII  (\definedterm{dB} \circ \definedterm{fst}) \) and \(\notationII  (\definedterm{dB} \circ \definedterm{snd}) \), or
equivalently, \(\notationII  (\isalambda\ (\localvar{x},\localvar{y}). \definedterm{dB}\ \localvar{x}) \) and \(\notationII  (\isalambda\ (\localvar{x},\localvar{y}). \definedterm{dB}\ \localvar{y}) \).

\begin{Lem}
\formal
\label{lem:hybrid-abstr2-is-componentwise}
\(\notationII  \definedterm{abstr\_2}\ \localvar{S} \isaholiff %
     \bigl( \bigl(  \isaholuniv\ \localvar{y}. \definedterm{abstr}\ (\isalambda\ \localvar{x}. \localvar{S}\ (\localvar{x}, \localvar{y})) \bigr) \isaholconj %
            \bigl(  \isaholuniv\ \localvar{x}. \definedterm{abstr}\ (\isalambda\ \localvar{y}. \localvar{S}\ (\localvar{x}, \localvar{y})) \bigr) \bigr) \)
\end{Lem}

This lemma shows that if a two-argument function satisfies \(\notationII \definedterm{abstr}\) in
each argument for any fixed value of the other argument, then it satisfies
\(\notationII \definedterm{abstr\_2}\).  (And the converse, which is easier.)
We omit the formal proof, but note that it is fairly long and requires
several lemmas.

Having thus reduced \(\notationII \definedterm{abstr\_2}\) to componentwise \(\notationII \definedterm{abstr}\),
we may now derive the desired simplification rule for the case
\(\notationII  (\definedterm{abstr}\ (\isalambda\ \localvar{x}. \definedterm{LAM}\ \localvar{y}. \localvar{W}\ \localvar{x}\ \localvar{y})) \).

\begin{Lem}
\label{lem:hybrid-abstr-LAM} \formal
\(\notationII  \negquad \isatheorem{abstr\_LAM}\colon \, 
    \isaholuniv\ \localvar{x}. \definedterm{abstr}\ (\isalambda\ \localvar{y}. \localvar{W}\ \localvar{x}\ \localvar{y}) \isaimpl \)\par\nopagebreak
\(\notationII  \phantom{ \negquad \isatheorem{abstr\_LAM}\colon \, } \quad %
     \definedterm{abstr}\ (\isalambda\ \localvar{x}. \definedterm{LAM}\ \localvar{y}. \localvar{W}\ \localvar{x}\ \localvar{y}) \isaholiff %
       ( \isaholuniv\ \localvar{y}. \definedterm{abstr}\ (\isalambda\ \localvar{x}. \localvar{W}\ \localvar{x}\ \localvar{y})) \)
\end{Lem}

This lemma provides a compositional rule for proving \(\notationII \definedterm{abstr}\) conditions
on functions of the form \(\notationII  (\isalambda\ \localvar{x}. \definedterm{LAM}\ \localvar{y}. \localvar{W}\ \localvar{x}\ \localvar{y}) \), via the reverse
direction of the biconditional.  Both directions are also used in the proof
of adequacy.  It was proved with the help of (a variant of)
\autoref{lem:hybrid-abstr2-is-componentwise}.%

We consider a small example, the term \(\notationII  (\definedterm{LAM}\ \localvar{x}. \definedterm{LAM}\ \localvar{y}. \localvar{x} \hybridapp \localvar{y}) \),
illustrating \(\notationII \isatheorem{abstr\_LAM}\) by proving that the argument of the outer
\(\notationII \definedterm{LAM}\) satisfies \(\notationII \definedterm{abstr}\), without the use of de~Bruijn syntax:
{\setbox0=\hbox{\(\notationII   \isaholuniv\ \localvar{x}. \definedterm{abstr}\ (\isalambda\ \localvar{x}. \definedterm{LAM}\ \localvar{y}. (\localvar{x} \hybridapp \localvar{y})) \qquad \)}
 \xdef\ajmtemp{\the\wd0}}
\begin{Display} \formal
\makebox[\ajmtemp][l]{\(\notationII   \isaholuniv\ \localvar{x}. \definedterm{abstr}\ (\isalambda\ \localvar{y}. \localvar{x}) \)}%
\(\notationII  \text{(by } \isatheorem{abstr\_const} \text{)} \)\par\nopagebreak
\makebox[\ajmtemp][l]{\(\notationII  \phantom{  \isaholuniv\ \localvar{x}. {}} \definedterm{abstr}\ (\isalambda\ \localvar{y}. \localvar{y}) \)}%
\(\notationII  \text{(by } \isatheorem{abstr\_id} \text{)} \)\par\nopagebreak
\makebox[\ajmtemp][l]{\(\notationII   \isaholuniv\ \localvar{x}. \definedterm{abstr}\ (\isalambda\ \localvar{y}. (\localvar{x} \hybridapp \localvar{y})) \)}%
\(\notationII  \text{(by } \isatheorem{abstr\_APP} \text{)} \)\par\nopagebreak\vspace{4pt}
\makebox[\ajmtemp][l]{\(\notationII  \phantom{  \isaholuniv\ \localvar{y}. {}} \definedterm{abstr}\ (\isalambda\ \localvar{x}. \localvar{x}) \)}%
\(\notationII  \text{(by } \isatheorem{abstr\_id} \text{)} \)\par\nopagebreak
\makebox[\ajmtemp][l]{\(\notationII   \isaholuniv\ \localvar{y}. \definedterm{abstr}\ (\isalambda\ \localvar{x}. \localvar{y}) \)}%
\(\notationII  \text{(by } \isatheorem{abstr\_const} \text{)} \)\par\nopagebreak
\makebox[\ajmtemp][l]{\(\notationII   \isaholuniv\ \localvar{y}. \definedterm{abstr}\ (\isalambda\ \localvar{x}. (\localvar{x} \hybridapp \localvar{y})) \)}%
\(\notationII  \text{(by } \isatheorem{abstr\_APP} \text{)} \)\par\nopagebreak\vspace{4pt}
\makebox[\ajmtemp][l]{\(\notationII  \phantom{  \isaholuniv\ \localvar{x}. {}} %
                         \definedterm{abstr}\ (\isalambda\ \localvar{x}. \definedterm{LAM}\ \localvar{y}. (\localvar{x} \hybridapp \localvar{y})) \)}%
\(\notationII  \text{(by } \isatheorem{abstr\_LAM} \text{)} \)
\end{Display}

Not only does the lemma \(\notationII \isatheorem{abstr\_LAM}\) allow \(\notationII \definedterm{abstr}\) statements
to be proved without the use of de~Bruijn syntax, but it also
completes the task of characterizing \(\notationII \isaholtype{expr}\) on its own terms~--
that is, without reference to the underlying de~Bruijn syntax.
This is demonstrated in \cite{martin:2010a} by the fact that representational
adequacy follows from Hybrid's lemmas concerning the type \(\notationII \isaholtype{expr}\),
and it is a significant improvement over both previous versions of Hybrid
\cite{ambler/crole/momigliano:2002,momigliano/martin/felty:2008}.

We also obtain the characterization of \(\notationII \definedterm{abstr}\) stated in
\abbrevref{sec:hybrid-overview} as a corollary of \(\notationII \isatheorem{abstr\_LAM}\):

\begin{Lem}
\label{lem:hybrid-expand-abstr} \formal
\(\notationII  \definedterm{abstr}\ \localvar{Y} \isaholiff \bigl( (\localvar{Y} = (\isalambda\ \localvar{x}. \localvar{x})) \isaholdisj {} \)\par\nopagebreak
\(\notationII  \phantom{\definedterm{abstr}\ \localvar{Y} \isaholiff \bigl(}
   ( \isaholexists\ \localvar{a}. \localvar{Y} = (\isalambda\ \localvar{x}. \definedterm{CON}\ \localvar{a})) \isaholdisj 
   ( \isaholexists\ \localvar{n}. \localvar{Y} = (\isalambda\ \localvar{x}. \definedterm{VAR}\ \localvar{n})) \isaholdisj {} \)\par\nopagebreak[3]
\(\notationII  \phantom{\definedterm{abstr}\ \localvar{Y} \isaholiff \bigl(}
   ( \isaholexists\ \localvar{S}\ \localvar{T}. \definedterm{abstr}\ \localvar{S} \isaholconj \definedterm{abstr}\ \localvar{T} \isaholconj 
     \localvar{Y} = (\isalambda\ \localvar{x}. \localvar{S}\ \localvar{x} \hybridapp \localvar{T}\ \localvar{x})) \isaholdisj {} \)\par\nopagebreak[3]
\(\notationII  \phantom{\definedterm{abstr}\ \localvar{Y} \isaholiff \bigl(}
   \bigl(  \isaholexists\ \localvar{W}. ( \isaholuniv\ \localvar{x}. \definedterm{abstr}\ (\isalambda\ \localvar{y}. \localvar{W}\ \localvar{x}\ \localvar{y})) \isaholconj 
   ( \isaholuniv\ \localvar{y}. \definedterm{abstr}\ (\isalambda\ \localvar{x}. \localvar{W}\ \localvar{x}\ \localvar{y})) \isaholconj {} \)\par\nopagebreak
\(\notationII  \phantom{\definedterm{abstr}\ \localvar{Y} \isaholiff \bigl( ( \isaholexists\ \localvar{W}. {}}
     \localvar{Y} = (\isalambda\ \localvar{x}. \definedterm{LAM}\ \localvar{y}. \localvar{W}\ \localvar{x}\ \localvar{y}) \bigr) \isaholdisj 
   (\localvar{Y} = (\isalambda\ \localvar{x}. \definedterm{ERR})) \bigr) \)
\end{Lem}

\section{Conclusion}
\label{sec:conclude}

Hybrid is the first approach to formalizing variable-binding
constructs that is both based on full HOAS and is built definitionally
in a general-purpose proof assistant (Isabelle\slash HOL).  More
recently, Popescu et. al. have developed an approach motivated by a
new proof of strong normalization for System F that takes advantage of
HOAS techniques~\cite{popescu/gunter/osborn:2010}.  It is also
definitional, implements full HOAS, and is implemented in Isabelle/HOL,
though the details of the formalizations as well as the case studies
carried out in each system are quite different.  A more
in-depth comparison is the subject of future work.

There are many other related approaches, and we mention only a few
here. See~\cite{felty/momigliano:2008, martin:2010a} for a fuller
discussion.  Systems that implement logics designed specifically for
reasoning using HOAS include Twelf~\cite{pfenning/schurmann:1999} (one
of the most mature systems in this category),
Abella~\cite{gacek:2008}, and Beluga~\cite{pientka/dunfield:2010}.
These systems have the advantage of being purpose-built for reasoning
about formal systems, but this can also be a disadvantage in that they
cannot exploit the extensive libraries of formalized mathematics
available for proof assistants such as Isabelle\slash HOL.
For a comparison of Hybrid to Twelf and Beluga,
see~\cite{felty/pientka:ITP2011}.
The nominal datatype package~\cite{Urban:JAR2008} implements a
different approach which seeks to formalize equivalence of classes of
terms up to renaming of bound variables, and also the Barendregt
variable convention, using concepts from nominal
logic~\cite{GabbayPitts:FAC2002,Pitts:IandC2003}.

There are several versions of Hybrid based on the Coq proof assistant.
One such version \cite{felty/momigliano:2008} closely follows the
structure of the Isabelle\slash HOL version; another implements a
constructive variant of Hybrid for Coq~\cite{capretta/felty:2006} that
aims to leverage the use of dependent types to simplify and provide
new ways to specify OLs.  There have also been a number of
applications and case studies for Hybrid, the largest being the
comparison of five formalizations of subject reduction for Mini-ML
with references~\cite{martin:2010a}, which uses
the improved Hybrid described in this paper.  Future work includes
porting other applications to use the new Hybrid.  This will be
straightforward since they are simpler and will be further simplified
by the new interface.  Future work also includes carrying out new
case studies to further illustrate the benefits of the new Hybrid.

Although we have significantly improved Hybrid, there is always room
for further improvement.  For example, the induction principle
discussed at the end of Sect.~\ref{sec:hybrid-overview} (the one whose
\(\notationII \definedterm{LAM}\) case is displayed) falls back to
named (or numbered) variables for inductive proofs, which means giving
up some of the advantages of HOAS.  We are working on a more general
approach to induction that preserves the HOAS feature of substitution
by function application.  In fact, we have proved an induction
principle for a type that represents $n$-ary functions on the type
\(\notationII \isaholtype{expr}\)~\cite{martin:2010a}, which we hope will serve as the basis
for general induction principles for HOAS in Hybrid.  Its integration
into Hybrid remains as future work.  As another example, we mentioned
that Hybrid is untyped, requiring predicates to be introduced to
distinguish different kinds of OL terms encoded into \(\notationII \isaholtype{expr}\).  On
one hand, these well-formedness predicates can provide a convenient
form of induction within the context of the two-level approach; on the
other hand this is a potential area for improvement.  Some work in
this direction has been done in the Coq version of
Hybrid~\cite{capretta/felty:2009}.


\end{document}